\documentclass[twocolumn,superscriptaddress,preprintnumbers,amsmath,amssymb,floatfix]{revtex4-1}
\pdfoutput=1
\usepackage{graphicx} % Include figure files
\usepackage{dcolumn} % Align table columns on decimal point
\usepackage{bm}% bold math
\usepackage{color}% color words
\usepackage{hyperref}% add hypertext capabilities
\usepackage{graphicx}
\usepackage{amsmath}
\usepackage{setspace}
\usepackage{appendix}

\begin{document}
\title{A Neural Network Approach for Orienting Heavy-Ion Collision Events}

\author{Zu-Xing Yang}
\affiliation{RIKEN Nishina Center, Wako, Saitama 351-0198, Japan}
\affiliation{School of Physical Science and Technology, Southwest University, Chongqing 400715, China}

\author{Xiao-Hua Fan}
\email{fanxiaohua@swu.edu.cn}
\affiliation{School of Physical Science and Technology, Southwest University, Chongqing 400715, China}
\affiliation{RIKEN Nishina Center, Wako, Saitama 351-0198, Japan}

\author{Zhi-Pan Li}
\affiliation{School of Physical Science and Technology, Southwest University, Chongqing 400715, China}

\author{Shunji Nishimura}
\affiliation{RIKEN Nishina Center, Wako, Saitama 351-0198, Japan}

\begin{abstract}

A convolutional neural network-based classifier is elaborated to retrace the initial orientation of deformed nucleus-nucleus collisions by integrating multiple typical experimental observables.
The isospin-dependent Boltzmann-Uehling-Uhlenbeck transport model is employed to generate data for random orientations of ultra-central uranium-uranium collisions at $E_\text{beam} =  1\,
 \text{GeV/nucleon}$.
% Given the existence of pronounced fluctuations and the impact of spectators in specific scenarios, the classifier still encounters a non-neglectable accuracy loss.
Statistically, the data-driven polarization scheme is essentially accomplished via the classifier, whose distinct categories filter out specific orientation-biased collision events.
This will advance the deformed nucleus-based studies on nuclear symmetry energy, neutron skin, etc.

\end{abstract}

\maketitle

\section{Introduction}

The multipole deformation of atomic nuclei is commonly attributed to collective motion arising from the interactions among valence nucleons \cite{Bohr1970PhysicsToday23.5860, Ring1980., Moeller2016At.DataNucl.DataTables109110.1204, Heyde2011Rev.Mod.Phys.83.1467}.
Traditionally, the quadrupole deformation can be measured experimentally from the rotational spectra of the nuclear excited state or the electric quadrupole moments derived from the hyperfine splitting of the atomic spectral line \cite{Moeller2016At.DataNucl.DataTables109110.1204}.
Since the year 2000, research on deformed nuclei through heavy-ion collisions has been initiated in the uranium-uranium collision system \cite{Li2000Phys.Rev.C61.021903}. 
Recently, at ultra-relativistic energies, a strong linear correlation between various-order anisotropic flows and multipole deformations was discovered \cite{Zhang2022Phys.Rev.Lett.128.022301, Jia2022Phys.Rev.C105.014905}, leading to significant implications for finding signatures of the nuclear deformation.

On the other hand, with the advancement of technology, machine learning has emerged as a powerful tool for investigating various properties of nuclear structure and reactions, particularly with respect to nuclear masses \cite{Niu2018Phys.Lett.B778.4853, Utama2016J.Phys.GNucl.Part.Phys.43.114002, Neufcourt2019Phys.Rev.Lett.122.062502, Athanassopoulos2004Nucl.Phys.A743.222235} and nuclear radii \cite{Yang2023Phys.Lett.B840.137870, Wu2020Phys.Rev.C102.054323, Dong2022Phys.Rev.C105.014308, Cotextquotesingle2022Phys.Rev.C105.034320}.
Aiming at nuclear deformation, deep neural network-based generative models are being developed for producing potential energy surfaces, rotational inertia, and vibrational inertia of deformed nuclei, which can be further utilized to investigate excitation spectra within the five-dimensional collective Hamiltonian approach \cite{Lasseri2020Phys.Rev.Lett.124.162502, Akkoyun2013PhysicsofParticlesandNucleiLetters10.528534}.
At the same time, a series of machine learning approaches, including support vector machine, artificial neural network \cite{Rumelhart1986Nature323.533536}, convolutional neural network (CNN) \cite{Bouvrie2006NotesOC}, and light gradient boosting machine \cite{Ke2017LightGBMAH}, were employed to determine impact parameters in heavy-ion collisions based on the simulation via quantum molecular dynamics (QMD) \cite{Kvasiuk2020J.HighEnergyPhys.2020., Haddad1997Phys.Rev.C55.13711375} and ultra-relativistic QMD (UrQMD) models \cite{Deng2022Phys.Lett.B835.137560, Saito2021Eur.Phys.J.A57., Kundu2021Phys.Rev.C104.024907}, achieving impact parameter identification accuracy within 0.8 fm \cite{Bass1996Phys.Rev.C53.23582363, Bass1994J.Phys.GNucl.Part.Phys.20.L21L26,Kuttan2020Phys.Lett.B811.135872,Li2021Phys.Rev.C104.034608}.
Subsequently, the 2D distributions of protons and neutrons on transverse momentum and rapidity plane are input to find signatures of the nuclear symmetry energy \cite{Wang2021Phys.Lett.B822.136669}, providing a reliable reference for experimental studies.

From the current research perspective, an important aspect missing in the studies about deformed nuclear reactions should be noted that, while the statistical effects of deformed nuclei on heavy-ion collisions have been revealed, the orientation issue regarding individual events has not yet garnered substantial focus. 
Absolutely or partially orienting the collision events will undoubtedly provide a wealth of information on the reactions and structure of deformed nuclei. 
For example, determining the symmetry dependence of collective flows in orientations without spectators would improve the description of the equation of state, which benefits nuclear astrophysics, such as binary neutron star merger simulations \cite{Huang2022Phys.Rev.Lett.129.181101}.
Moreover, exploring higher-order deformation effects through specific orientations would yield valuable insights.

This work aims to develop a multi-input CNN to map the nuclear initial state orientations of the uranium-uranium reaction systems. 
Since experimental orientation information is lacking, the network will be initially trained using simulated results based on the isospin-dependent Boltzmann-Uehling-Uhlenbeck (IBUU) transport model \cite{Li2004Phys.Rev.C69.011603,Li2005Phys.Rev.C71.014608}. 
Once trained with statistical observables matching experimental data reasonably well, the network can further filter the orientations for experiments.

\section{the isospin-dependent Boltzmann-Uehling-Uhlenbeck transport model \label{sec2}}

The IBUU transport model employs the Monte Carlo method to simulate the phase-space evolution of baryons and mesons during heavy-ion collisions, encompassing essential physical processes such as elastic and inelastic scattering, particle absorption, and decay \cite{Bertsch1988Phys.Rep.160.189233}.
The version we employed \cite{Yang2021J.Phys.GNucl.Part.Phys.48.105105, Yong2016Phys.Rev.C93.014602, Yang2018Phys.Rev.C98.014623,Guo2019Phys.Rev.C100.014617,Cheng2016Phys.Rev.C94.064621} has incorporated the Coulomb effect \cite{Yang2018Phys.Rev.C98.014623}, Pauli blocking, and medium effects on scattering cross sections \cite{Xu2011Phys.Rev.C84.064603}, etc.  
The used single nucleon potential includes a Skyrme-type parametrization isoscalar term and an exponential isovector term \cite{Yong2016Phys.Rev.C93.014602,Cherevko2014Phys.Rev.C89.014618}, which reads
\begin{equation}
    U(\rho) = A (\rho / \rho_0) + B (\rho / \rho_0)^\sigma,
\end{equation}
where $\sigma = 1.3$,  $A = -232\,\text{MeV}$  accounts for the attractive part, and $B = 179\,\text{MeV}$ accounts for the repulsive part. 
These choices correspond to an incompressibility coefficient $K = 230\,\text{MeV}$.

\begin{figure}[tb]
\includegraphics[width=8.5 cm]{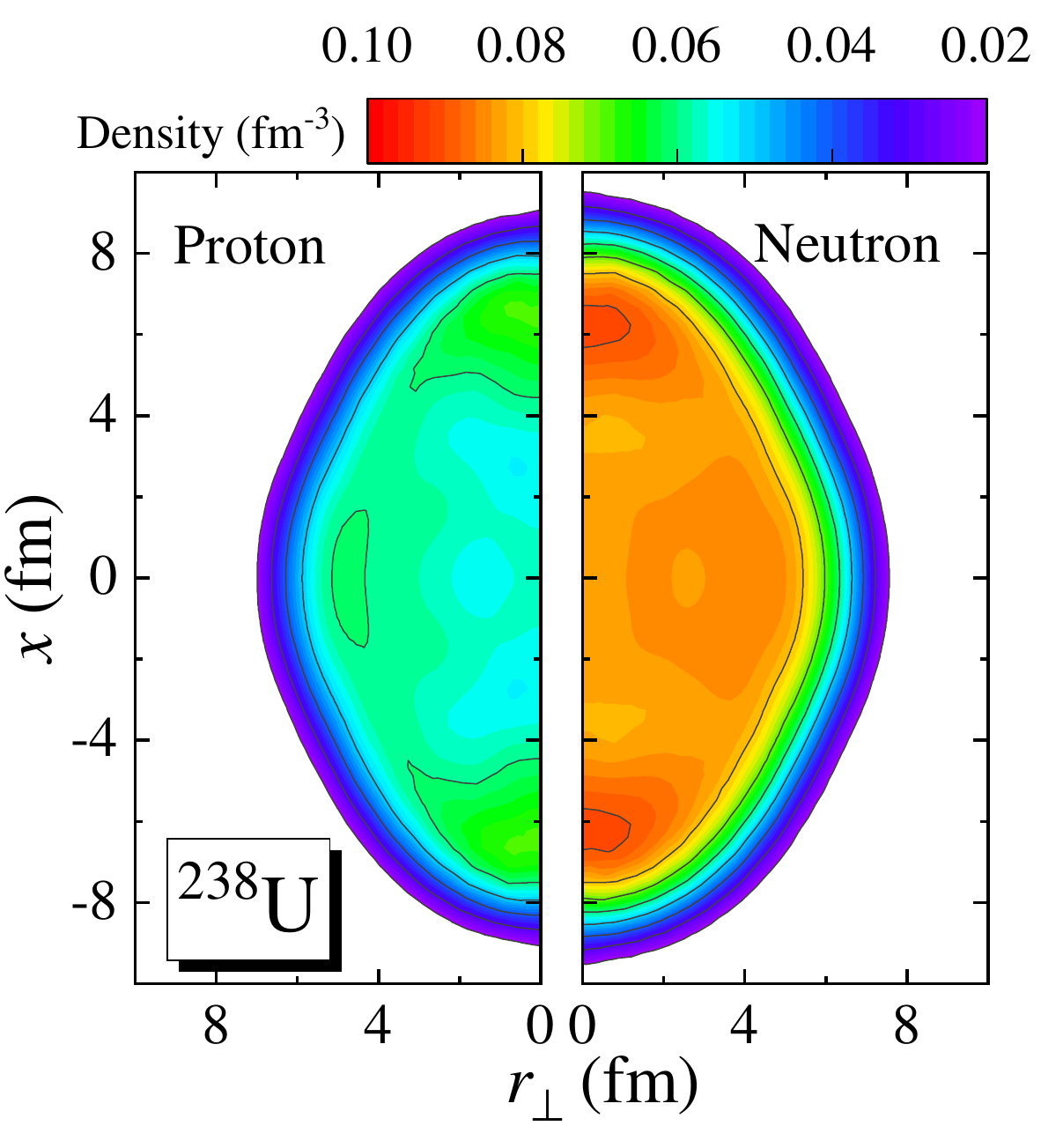}
\caption{\label{fig1} The ground-state proton (left panel) and neutron (right panel) densities for $^{238}\text{U}$ in the $r_\bot-x$ ($r_\bot = \sqrt{(y^2 + z^2)}$) coordinate system calculated by the relativistic mean-field model plus a Bardeen-Cooper-Schrieffer pairing.
  }
\end{figure}

The nucleon density distributions for $^{238}$U are calculated by the relativistic mean-field model plus a Bardeen-Cooper-Schrieffer pairing (RMF+BCS) \cite{GAMBHIR1993Mod.Phys.Lett.A08.787795, Li2022Phys.Rev.C106.024307}. 
The point-coupling PC-PK1 functional \cite{Zhao2010Phys.Rev.C82.054319} is used in the particle-hole channel and a separable pairing force is for the particle-particle channel.
Due to the superior capability in delineating the nuclear radii, profiles, and deformations for neutron-rich nuclei, the densities generated by RMF models have recently been utilized to initialize the IBUU model \cite{Fan2023Phys.Rev.C108.034607, Fan2019Phys.Rev.C99.041601}.
Shown in Fig.~\ref{fig1} is the nucleon density distributions of $^{238}\text{U}$ with the quadrupole deformation being $\beta=0.287$.

\begin{figure}[tb]
\includegraphics[width=8.5 cm]{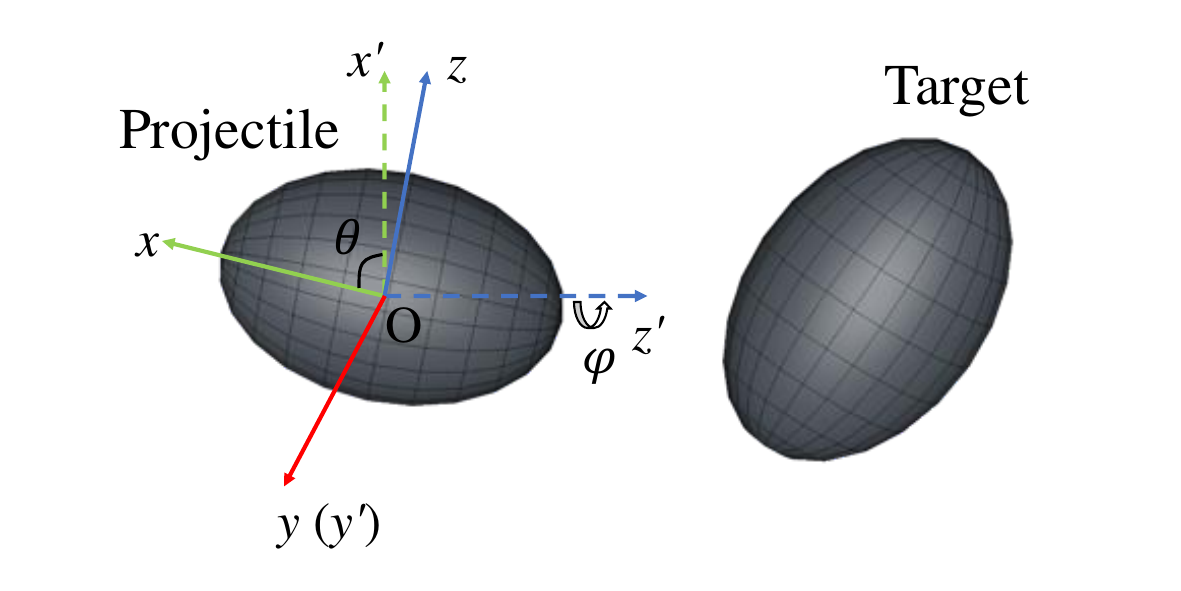}
\caption{\label{fig2} Simulated schematic for uranium-uranium collision, where $x-z$ plane is the reaction plane. 
  }
\end{figure}

Initially, the intrinsic coordinate systems ($xyz$) of the colliding nuclei are set to align with the center-of-mass coordinate system ($x'y'z'$) of the reaction.
As shown in Fig.~\ref{fig2}, an Euler rotation operator ($\Omega(\varphi,\theta,0) = R_z(\varphi)R_y(\theta)R_x(0)$) is applied independently to the target and the projectile, where only four degrees of freedom ($\theta_1,\varphi_1$ for the target, and $\theta_2,\varphi_2$ for the projectile) are required for each collision event due to the absence of triaxial deformation.
Considering that spectator fragments from the projectile can be detected on an event-by-event basis in experiments \cite{Foehr2011Phys.Rev.C84.054605}, we confine the study to the events where the projectiles are completely obstructed by the targets as
\begin{equation}
    \theta_2 \in 
\begin{cases}
    [\theta_1, 180^\circ-\theta_1] &  \theta_1 < 90^\circ\\
    [180^\circ-\theta_1, \theta_1] &  \theta_1 \geq 90^\circ
\end{cases}
\end{equation}
and 
\begin{equation}
    \varphi_2 = \varphi_1
\end{equation}
with $\theta_1 \in [0^\circ, 180^\circ]$ and $\varphi_1 \in [0^\circ, 180^\circ]$.
Given the limited discriminative power of impact parameters, the scenario of ultra-central collisions is set with $b \leq 1$ fm.
In experiments, events with the target spectator fragments appearing only in one direction can also be filtered out to determine ultra-central collisions.
Under the present densities, the Fermi gas model with local density approximation is employed to generate the momentum distributions.
Within the intermediate energy range, increasing the beam energy will enhance the deformation effects \cite{Fan2023Phys.Rev.C108.034607}, and also will produce more hadrons within a single event, which is advantageous to trace back the orientation. 
Accordingly, the beam energy is set to $1\,\text{GeV/nucleon}$.

In the classic picture, the orientation of collisions is random, necessitating averaging over all events to search for the signatures of deformation.
For each event, BUU-type models typically utilize $N$ test particles to mimic the spreading of real nucleons in phase space, thereby overcoming the limitations of point particle simulations \cite{Li2023Nucl.Phys.A1039.122726}, which means, that every $N$ collisions share a common mean field evolved from the same impact parameters and collision orientations \cite{Bertsch1988Phys.Rep.160.189233}.
This also implies that the $N$ determines the stability of the mean field and defines the fluctuations in observables for each event.
In this study, determining the collision orientation requires ensuring that the observable fluctuations for theoretical simulation events closely mirror those observed in experiments.
To this end, we vary the test particle number from $N=1$ to $N=100$ to simulate a real event, where $N=1$ signifies the dynamics evolution of point particles, while larger $N$ values can be understood as particles converging towards a wave packet.
On this basis, collisions with $10,000$ different collision orientations are simulated, and the data is split into training and validation sets in a 7:3 ratio.

\begin{figure}[tb]
\includegraphics[width=8.5 cm]{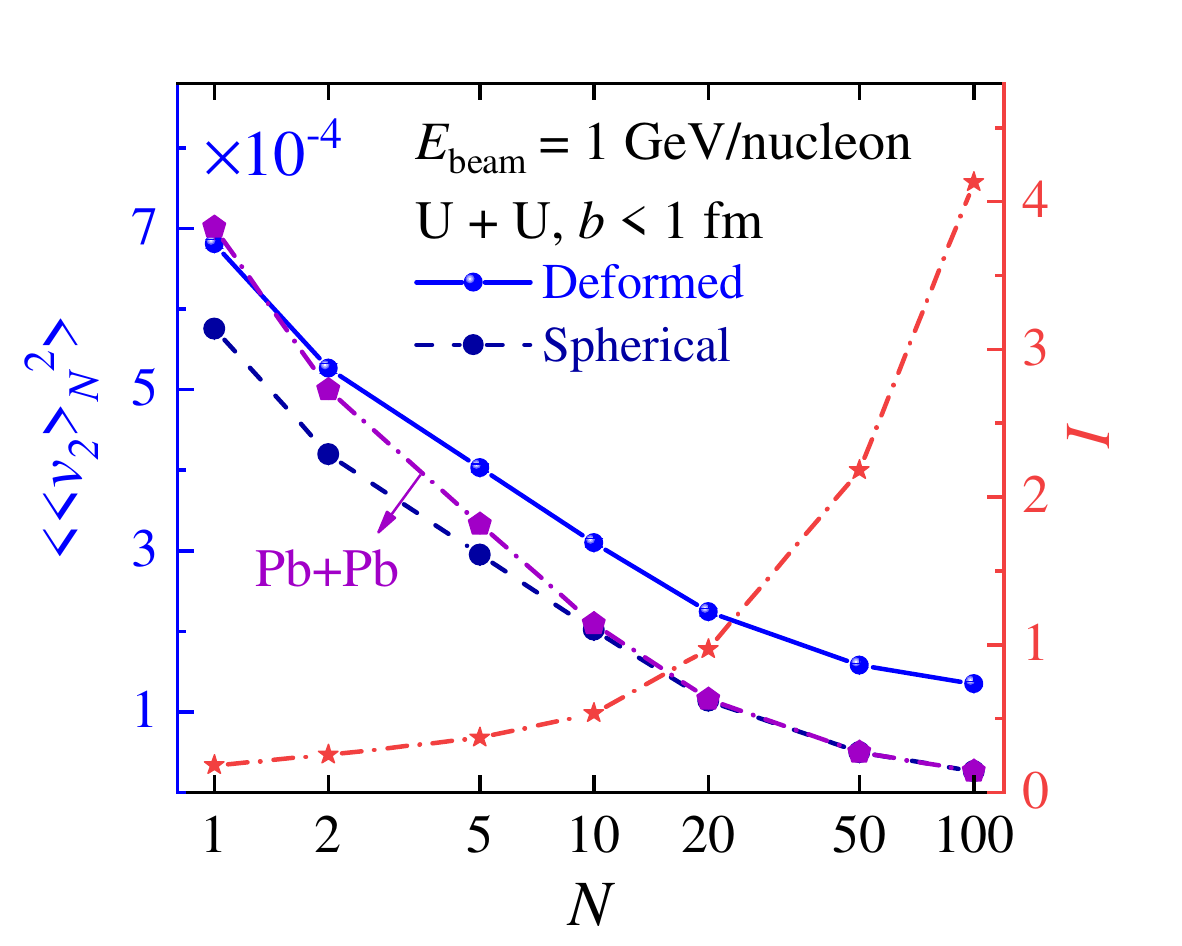}
\caption{\label{fig4} The mean square elliptic flows as a function of the test particle number $N$ (curves in the blue color series) and the corresponding strength of the deformation effect $I$ (curves in the red color).
The purple curve serves as a contrast, representing the spherical Pb+Pb collision system.
  }
\end{figure}

In Fig.~\ref{fig4}, the observable $\langle \langle v_2 \rangle_N^2 \rangle$  is examined (the blue color series), as it has been proven to be a sensitive probe of quadrupole deformation \cite{Jia2022Phys.Rev.C105.014905}.
The inner and outer brackets of $\langle \langle v_2 \rangle_N^2 \rangle$ respectively represent the averaging elliptic flow over $N$ test-particle collisions with the same orientation and averaging over all orientations.
The response relation between $\langle \langle v_2 \rangle_N^2 \rangle$ and the quadrupole deformation $\beta$ can be expressed as
\begin{equation}
    \langle \langle v_2 \rangle_N^2 \rangle = a_N + f(\beta),
\end{equation}
where the fluctuation $a_N = \langle \langle v_2 \rangle_N^2 \rangle_{\beta = 0}$ corresponds to the blue dashed curve and the deformation effect $f(\beta)$ is manifested as the difference between the blue solid and dashed curves.
In this case, we define the strength of deformation effect as 
\begin{equation}
    I=\frac{f(\beta)}{a_N},
\end{equation}
which corresponds to the red curve of Fig.~\ref{fig4}.
The significant effect on the mean square elliptic flow indicates the feasibility of exploring nuclear deformations via intermediate-energy heavy-ion collisions.
It can be noted that the increasing $N$ leads to a reduced fluctuation, while the $f(\beta)$ remains constant during this process, resulting in increasing strength of deformation effect.
In comparing the spherical collision systems of Pb+Pb, we observe that increasing $N$ amplifies the differences between the two systems, and this difference saturates at $N=20$. 
These properties, especially the ratio and difference between the two systems, can be exploited to reference experiments to determine reasonably the $N$.
Generally, $N$ is set to be at least 50 to maintain the stability of the mean field \cite{Li2023Nucl.Phys.A1039.122726}, thus this study will also be going forward based on this premise.

\section{Convolutional Orientation Filter \label{sec3}}

\begin{figure*}
\centering
\includegraphics[width=14 cm]{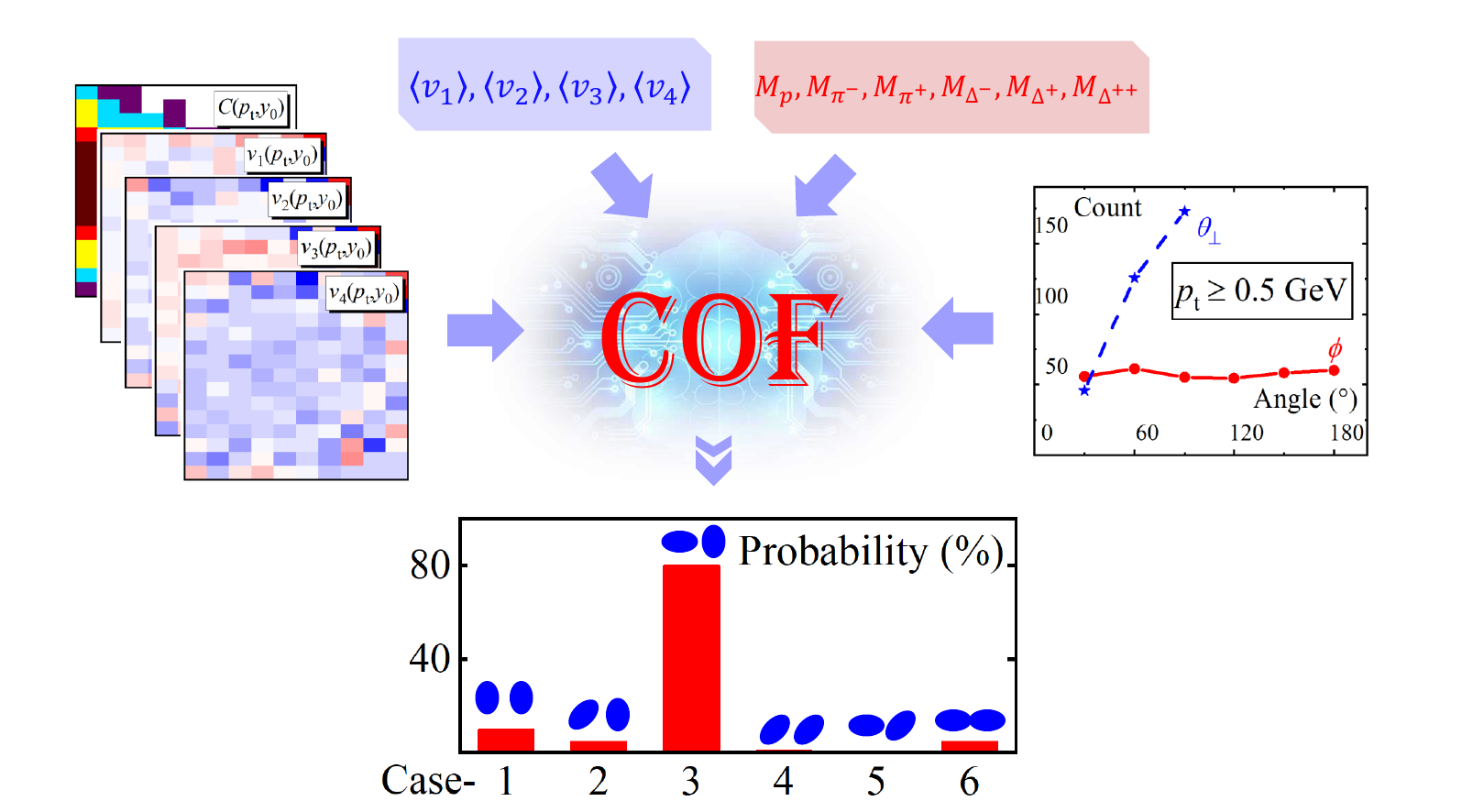}
\caption{\label{fig3} Schematic diagram of the structure of the convolutional orientation filter (COF) neural network.
  }
\end{figure*}

In the IBUU transport model, the comprehensive information of the final state is stored in the phase-space distribution $f(\vec{r}, \vec{p}, T)$, where $\vec{r}$ and $\vec{p}$ represent the coordinate and momentum of a hadron, respectively. 
The variable $T$ denotes the type of the hadron, which includes $p$, $n$, $\pi^+$, $\pi^0$, $\pi^-$, $\Delta^{++}$, $\Delta^{+}$, $\Delta^{0}$, $\Delta^{-}$.
Regarding the reaction picture at the current energy, there has been ample discussion in recent work on the principal component analysis of azimuthal flow \cite{Li2023Nucl.Phys.A1034.122640}.
We present a summary of empirically considered observables pertaining to collision orientation as follows:

\begin{enumerate}
  \item The hadron counting $C(p_t,y_0)$ and distributions of anisotropic flows $v_{n=1,2,3,4}(p_t,y_0)$ in central rapidity $y_0$($=y/y_\text{beam}$) and transverse momentum $p_t$ ($=\sqrt{p_x^2+p_y^2}$). 
  In practice, we adopt grid intervals of $\Delta p_t = 0.15\,\text{GeV/c}$ and $\Delta y_0 = 0.1$ for transverse momentum and center rapidity bins, respectively.
  \item Mean values of anisotropic flows $\left\langle v_{n} \right\rangle (n=1,2,3,4)$ for all the emitted particles. 
  \item Multiplicities of charged particles $M_p$, $M_{\pi^-}$, $M_{\pi^+}$, $M_{\Delta^-}$, $M_{\Delta^+}$, $M_{\Delta^{++}}$, where both medium and free particles are incorporated.   
  \item Counting of the high-momentum nucleons ($p_t > 0.5 \,\text{GeV/c}$) in transverse emission azimuth angle $\phi$ ($=\arccos{(p_x/p_t)}$) and longitudinal emission angle $\theta_{\bot}$ ($=\arccos{(p_z/|\Vec{p}|)}$). The angular intervals are uniformly set at $\Delta\theta_{\bot}(\Delta\phi)=30^\circ$.
\end{enumerate}
In particular, the anisotropic flows are expressed as
\begin{equation}
    v_n = \left\langle\cos(n\phi) \right\rangle.
\end{equation}
The organized observables will serve as inputs for the network, as depicted in Fig.~\ref{fig3}.
Hereinafter, the network will be denoted as the convolutional orientation filter (COF).
The output was anticipated to include $\theta_1$, $\theta_2$, $\varphi_1$, and $\varphi_2$, nevertheless, after multiple rounds of training and validation, we have arrived at the following conclusions:
1.~According to the Heisenberg uncertainty principle, considering this issue as a linear regression task is relatively less reliable compared to adopting a classification approach;
2.~The angle $\varphi$ demonstrates limited sensitivity to the present inputs;
3.~$\theta = k$ and $\theta = 180^\circ-k$ are degenerate states and cannot be distinguished.
As such, we categorize the output, based on the $\theta_1$ and $\theta_2$, into six cases (as illustrated in Table~\ref{tab1}) and perform one-hot encoding for representation.
In this case, a tip-body collision ($\theta_1 = 0^\circ, \theta_2 = 90^\circ$) would be labeled as $[0, 0, 1, 0, 0, 0]$.
\begin{table}[h]
\centering
\caption{\label{tab1} The initial collision orientations corresponding to the six classification cases.}
\renewcommand{\arraystretch}{1.2}
\begin{tabular}{c|cc}
\hline\hline
  Case   & ~~~$\theta_1$ (Target)~~~           & ~~$\theta_2$ (Projectile)~~       \\ \hline
1   & $[0^\circ,30^\circ]\cup[150^\circ,180^\circ]$ & $[0^\circ,30^\circ]\cup[150^\circ,180^\circ]$ \\ 
2   & $[0^\circ,30^\circ]\cup[150^\circ,180^\circ]$ & $[30^\circ,60^\circ]\cup[120^\circ,150^\circ]$ \\ 
3   & $[0^\circ,30^\circ]\cup[150^\circ,180^\circ]$ & $[60^\circ,120^\circ]$ \\ 
4   & $[30^\circ,60^\circ]\cup[120^\circ,150^\circ]$ & $[30^\circ,60^\circ]\cup[120^\circ,150^\circ]$ \\ 
5   & $[30^\circ,60^\circ]\cup[120^\circ,150^\circ]$ & $[60^\circ,120^\circ]$ \\ 
6   & $[60^\circ,120^\circ]$ & $[60^\circ,120^\circ]$ \\ \hline\hline
\end{tabular}
\end{table}

The four aforementioned inputs are individually fed into multiple branch cells composed of convolutional, pooling, and fully connected layers.
The outputs of these branch cells are then aggregated in a trunk cell and subjected to batch normalization \cite{Ioffe2015.}, ensuring a stable feature distribution. 
An essential point that must be emphasized is that the network consists of 15 layers with intricate connections, whose complexity can lead to some neurons being trapped in the negative range and becoming deactivated under the commonly used $\text{ReLU}$ \cite{Villani2023.} activation function for nonlinearity, especially after multiple iterations. 
To address this issue and improve convergence, we adopt the LReLU activation function \cite{Xu2015.} to avoid gradient vanishing.
Ultimately, after multiple iterations constrained by the cross-entropy loss function \cite{Mao2023.}, the COF network generates orientation classifications normalized by the Softmax function \cite{Wang2018.223226}.
The meticulously designed hyperparameter set for the COF network architecture is listed in Appendix~\ref{sec:a}, which can also be employed for other transport models, such as a multiphase transport model \cite{Bzdak2014Phys.Rev.Lett.113.252301, Ma2014Phys.Lett.B739.209213, Wolter2022Prog.Part.Nucl.Phys.125.103962}, UrQMD model \cite{Deng2022Phys.Lett.B835.137560, Saito2021Eur.Phys.J.A57., Kundu2021Phys.Rev.C104.024907}, and antisymmetrized molecular dynamics model \cite{Frosin2023Phys.Rev.C107.044614, Takatsu2023Phys.Rev.C107.024314}, etc.

\section{Results and analysis \label{sec:4}}

We shift our attention to the performance of the COF network, which is showcased in Table~\ref{tab2}.
The training set and validation set show only marginal distinctions, since the abundance of simulated events used for training has effectively mitigated overfitting, showcasing a strong generalization capability. 
This, to some extent, supports the application of the network to experimental data.
The accuracy demonstrates a positive correlation with $I$, while once $N$ surpasses 20, the enhancement in accuracy becomes quite limited.
It is understandable that some collision scenarios positioned at the boundary between two classes and with the random impact parameters interfered cannot be correctly classified.

\begin{table}[h]
\centering
\caption{\label{tab2} Accuracy of the training and validation sets with $N$ test particles to simulate a real event.}
\renewcommand{\arraystretch}{1.2}
\begin{tabular}{c|cccc}
\hline\hline
   $N$  &  ~~~~1~~~~      & ~~~~20~~~~ & ~~~~50~~~~  & ~~~~100~~~~   \\ \hline
training set   & 22.8\% & 67.5\% & 72.4\% & 77.1\% \\ 
validation set   & 22.0\% & 65.1\% & 72.2\% & 74.1\%\\ 
\hline\hline
\end{tabular}
\end{table}

\begin{figure}[tb]
\includegraphics[width=9 cm]{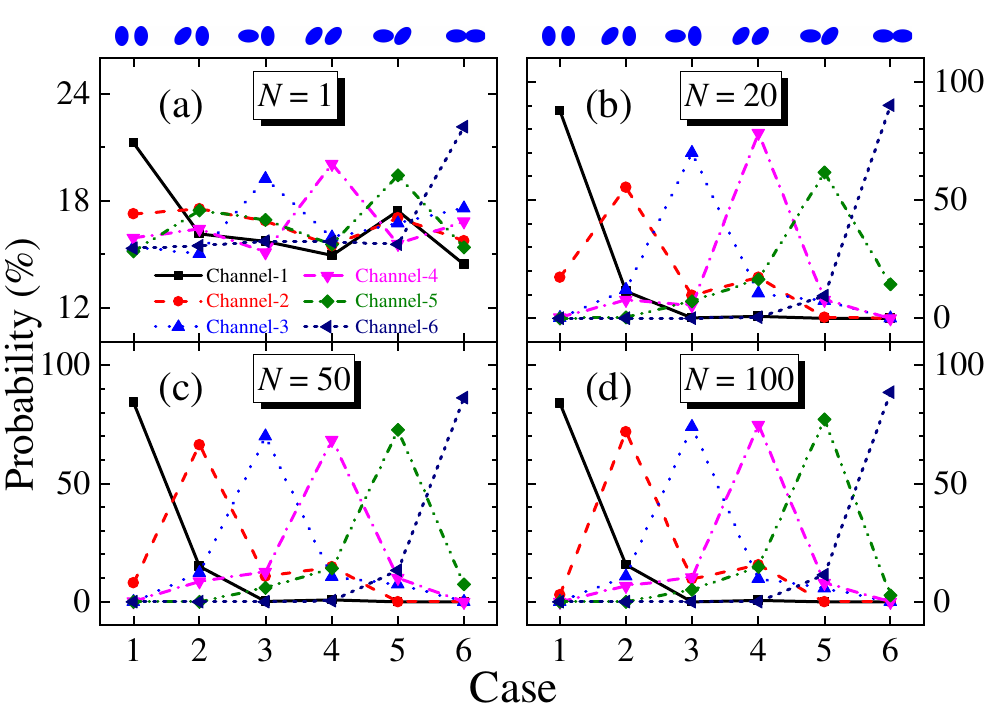}
\caption{\label{fig5} The probability distribution of the actual case at different predicted channels with (a) $N = 1$,  (b) $N = 20$, (c) $N = 50$, and  (d) $N = 100$.
  }
\end{figure}

Displayed in Fig.~\ref{fig5} are the probability distributions of the six classification cases at different output channels, where the normalization condition is $\sum_{i(j)} P_{i(j)} = 1$ with the subscript $i,j$ representing the output channels and the true cases, respectively.   
Taking the black solid line in (c) $N=50$ as an example, the collision events classified as Channal-1 are actually composed of $80\%$ Case-1 and $20\%$ Case-2.
For the descriptions of point particles in Fig.~\ref{fig5}(a), the network polarization performance is quite weak, only the partial polarization brought by Channel-1 and -6 remains effective.
As the $N$ increases to 20 in panel (b), the results become considerably reliable.
The situation with $N=50$ and $N=100$ in panels (c) and (d) provides enhanced descriptions primarily for categories with a higher number of spectators (Case-2, -3, and -5).
It is evident that Channel-1 and -6 still exhibit better discriminative power, which corresponds to the commonly mentioned body-body and tip-tip collisions \cite{Li2000Phys.Rev.C61.021903, Jia2022Phys.Rev.C105.014905} and are crucial for investigating various properties of deformed nuclei.
Moreover, leveraging the high angular resolution, conducting combined analyses across multiple channels would be beneficial. 
For example, by specifically choosing events categorized in Channel-1, -4, and -6, the spectator effects could be effectively filtered out. 
Alternatively, amalgamating data from Channel-1, -2, and -3 allows for the exclusion of events involving large-angle target nucleus ($\theta_2>30^\circ$).

\begin{figure*}[tb]
\includegraphics[width=18 cm]{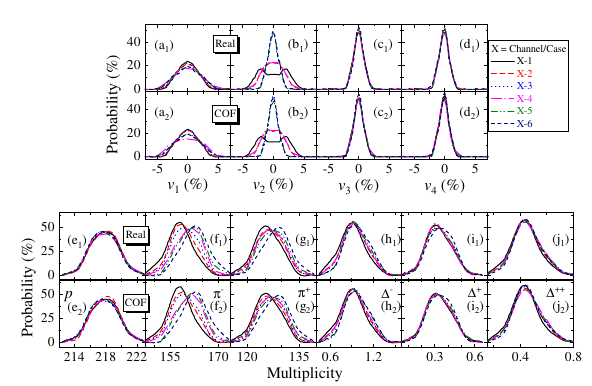}
\caption{\label{fig6} The distributions of event count proportions for the anisotropic flows $v_1$ (a$_x$), $v_2$ (b$_x$), $v_3$ (c$_x$), $v_4$ (d$_x$) and the multiplicities  of charged particle including $p$ (e$_x$), $\pi^-$ (f$_x$), $\pi^+$ (g$_x$), $\Delta^-$ (h$_x$),  $\Delta^+$ (i$_x$), $\Delta^{++}$ (j$_x$) with $N = 50$ on validation set, where subscript $x=1$ represents the actual classification cases, while $x=2$ represents the situation of the COF output channels.
  }
\end{figure*}

Two key issues have arisen: the extent to which observables play a role in the identification of orientations, and the impact of precision loss on the observables.
It is possible to make a judgment by comparing the COF classifications with the real scenarios, for which the count proportion distributions of various observables on the validation set are displayed in Fig.~\ref{fig6}.
The first two rows contrast the discrepancies between the COF classifications and the real scenarios for the anisotropic flows $v_1$, $v_2$, $v_3$, and $v_4$,  while the last two rows contrast the differences for charged particles $p$, $\pi^-$, $\pi^+$, $\Delta^-$, $\Delta^+$, $\Delta^{++}$.
It can be observed that the directed flow $v_1$, elliptic flow $v_2$, and multiplicities of $\pi^-$ and $\pi^+$ are more sensitive to orientation. 
This indicates that these observables play a more crucial role in the classification.
Comparative analysis of subscripts $x=1$ and $x=2$ in Fig.~\ref{fig6} reveals negligible differences between the COF classification and actual categories, even with a precision loss exceeding 20\%.
The successful classification of these untrained events emphasizes the effectiveness of the current network.
This also means that further improving accuracy is dispensable due to the extensive overlap in the distribution of observables across channels.
In fact, the width of these distributions is a direct manifestation of fluctuations, which has been verified to decrease with $N$ increasing gradually. 
A narrower distribution results in reduced overlap amongst different channels/cases, thereby further improving the accuracy.
From an experimental perspective, measuring the full width at half maximum of non-sensitive observables, such as $v_3$, and $v_4$, can also aid in determining a reliable value for $N$.

\begin{figure}[tb]
\includegraphics[width=7.5 cm]{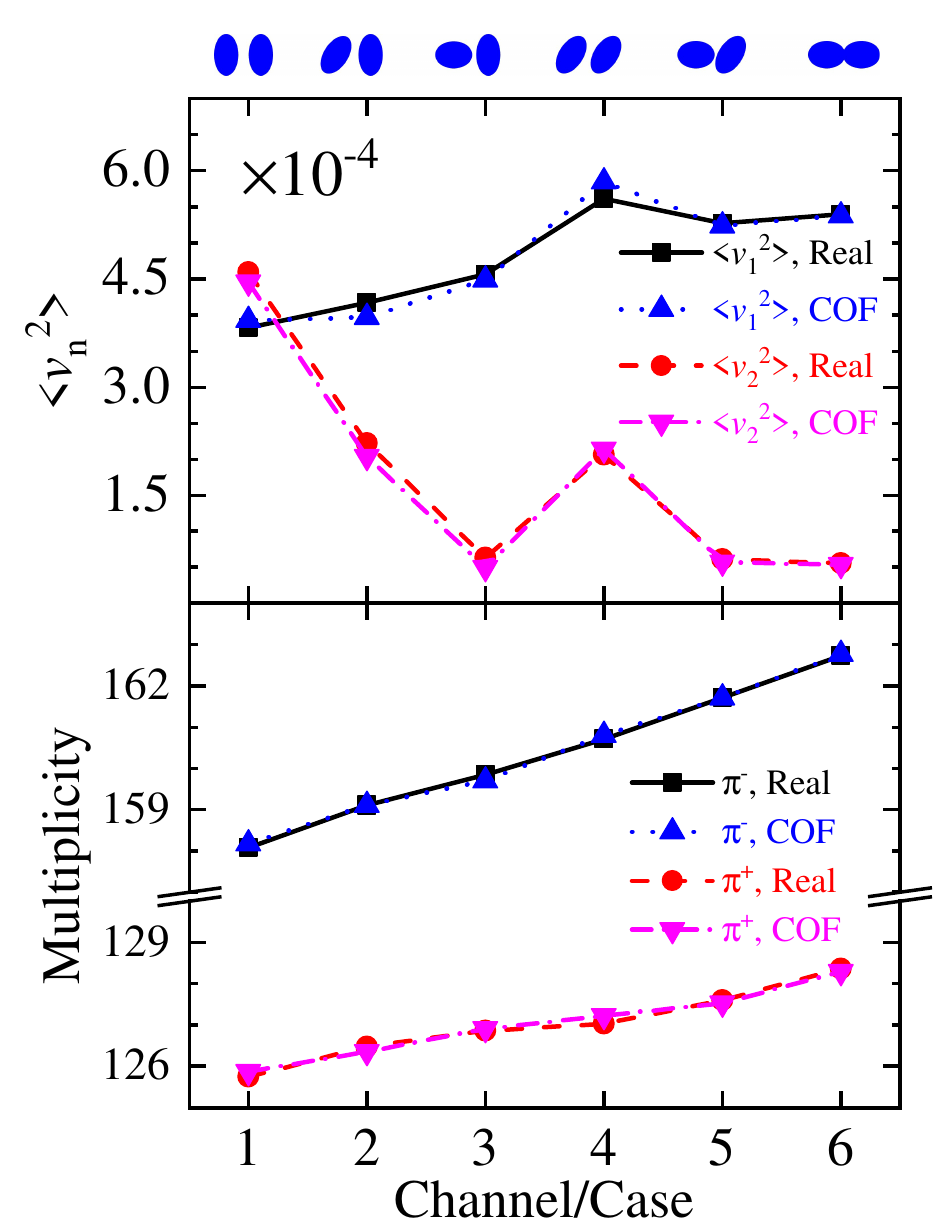}
\caption{\label{fig7} Upper panel: The distribution of mean squared directed flow $\langle  v_1^2 \rangle$ and elliptic flow $\langle  v_2^2 \rangle$  across different predicted channels. 
Lower panel: The distribution of $\pi^-$ and $\pi^+$ multiplicities across different predicted channels. 
  }
\end{figure}

The mean squared collective flows $\langle v_1^2 \rangle$, $\langle v_2^2 \rangle$, and the multiplicities of $\pi^-$ and $\pi^+$, as sensitive observables, are further investigated in Fig.~\ref{fig7}.
The results of the screened events align exactly with the theoretical calculation, affirming the accuracy of COF reaches a level where the accuracy loss is no longer sensitive to the observable.
Such statistical regularities emphasize a key point that it is challenging to distinguish the orientations of colliding nuclei through conventional data processing methods.
As such, the emerging artificial intelligence technologies geared toward combining multiple observables may be the only solution to this problem.

For $\langle v_1^2 \rangle$, the overall trend suggests that as the collision scenario approaches tip-tip configuration, resulting in higher central density, more emitting particles exhibit a bias toward the $x$-direction, which is the direction of the impact parameter. 
Interestingly, when both colliding nuclei are oriented at approximately a $45^\circ$ to the $z$-axis, particles appear to exhibit a greater tendency to be squeezed out in the $x$-direction.
In $\langle v_2^2 \rangle$, the six channels (cases) are observed to be statistically divided into three categories. 
Channel-1 exhibits a higher distribution in regions with stronger $v_2$, Channel-2 and -4, as the second tier, and Channel-3, -5, and -6 encompass the category of events with the weakest elliptic flow.
This order is clearly related to the initial state overlap eccentricity, where the overlap refers to the projection of the participants on the $x-y$ plane \cite{Bertsch1988Phys.Rep.160.189233}. 
Simultaneously, the substantial spectators in Case-2, -3, and -5 contribute to a partial suppression of their $\langle v_2^2 \rangle$.
In the combined influence of these factors, we find that these three categories appear to be independent of the target orientation and are solely determined by the projectile orientation.
In particular, Channel-1, representing body-body collisions, vividly amplifies the deformation effect, which will promote future research on nuclear structural properties and reaction mechanisms.
Turning to the lower panel of Fig.~\ref{fig7}, it can be noted that collisions approaching tip-tip orientation tend to enhance the yield of pion mesons, attributed to the higher reaction density achievable in tip-tip collisions.

During tracing the collision orientations, a multitude of nonlinear transformations interact, making it difficult to determine which input variable has a significant impact on the feed-forward process of the COF network. 
Aiming at the other inputs, including the 2D distributions of the hadron counting and anisotropic flows on transverse momentum and rapidity plane and the counting of the high-momentum nucleons in transverse (longitudinal) emission angle, it is challenging to discern the impact of orientation on an individual event from a conventional perspective.
Nevertheless, the contribution of these inputs to orientation recognition cannot be overlooked, which is analogous to the situation that using deep learning to find signatures of the nuclear symmetry energy \cite{Wang2021Phys.Lett.B822.136669}.
Eventually, these observables collectively facilitate the recognition of the collision orientation. 

With the aid of the COF network, orientation-dependent physical quantities would be experimentally measured, which will enable us to capture more information from facilities such as HIAF/IMP, J-PARC, FAIR, NICA, HIMAC, etc.
The previous experiments are also worth more detailed analysis, especially the Au+Au reactions at 200 GeV and the U+U reactions at 193 GeV at RHIC \cite{Adamczyk2015Phys.Rev.Lett.115.222301}.
In future research, utilizing the elliptic flow in Channel-1 might offer a geometric effect-based means to explore further the impact of symmetry energy and equation of state.
Furthermore, theoretical research \cite{Fan2022Phys.Lett.B834.137482} has indicated that high-momentum tails, resulting from short-range correlations, are more likely to manifest at the nuclear surface. 
Therefore deformed nuclei with larger surface areas will undoubtedly provide an excellent platform for investigating this phenomenon, which will also benefit from the COF network.
In summary, this research would greatly broaden the horizons of heavy-ion collision studies, facilitating a deeper comprehension of both nuclear structure and nuclear reactions.

\section{summary \label{sec:6}}

Based on the 1 GeV/nucleon uranium-uranium ultra-central collision data simulated using the IBUU transport model, we construct a composite neural network of multiple observables in a supervised classification mode to retrospectively determine the initial state orientation of the deformed colliding nuclei.
This data-driven polarization approach is applicable in a statistical sense.
The different output channels of the COF network filter out specific orientation-biased collision events.
By simulating the evolution of point particles ($N=1$) and the wave packet composed of multiple point test particles ($N=20,50,100$) in the mean field, we conclude that the identification capability of the COF network is positively correlated with the ratio of the deformation effect to fluctuation.

When the evolved results are composed of $N = 50$ test particles, the accuracy for both tip-tip and body-body collisions exceeds 80\%.
Upon comparing observables between real categories and COF channels, including anisotropic flows and charged particle multiplicities, we identify virtually no discernible differences, which is a reflection of the extensive overlap in the distribution of observables across channels.
Amongst these observables, we note the orientation dependence of $v_1$, $v_2$, and the pion multiplicities, leading to the conclusion that the eccentricity of the overlap, the achieved density in the reaction, and the spectator effect profoundly impact these sensitive measurements.

At relativistic energies, the production of hadrons is often increased by orders of magnitude compared to intermediate energies, while participants are no longer influenced by spectators, making the identification of the initial state orientation markedly more reliable.
We look forward to the COF network being combined with a broader range of transport models and applied to experimental data measurements.

\section{Acknowledgements}
We acknowledge helpful discussions with Prof.~Haozhao Liang. 
This work is supported by the National Natural Science Foundation of China under Grants Nos. 12005175, 11875225, 12375126,
the JSPS Grant-in-Aid for Scientific Research (S) under Grant No. 20H05648, 
and the RIKEN Projects: r-EMU, RiNA-NET, and the INT Program INT-23-1a and Institute for Nuclear Theory.
This work is also supported by the Fundamental Research Funds for the Central Universities under Grant No. SWU119076.

\appendix

\section{The Hyperparameter Set of Convolutional Orientation Filter \label{sec:a}}
  
\begin{table}[h]
\caption{\label{tab3}The hyperparameter set of the convolutional orientation filter structure. The ``$D$" represents the output dimension of the layer, the ``$C_\text{in}$" denotes the input channels, the ``$C_\text{out}$" signifies the output channels, the ``$K_s$" refers to the size of the kernel which includes both convolutional and pooling dimensions, the ``$S_t$" indicates the stride used during convolution or pooling operations, and the ``$g(x)$" represents the non-linear activation function. 
See the text for abbreviations.}
\renewcommand{\arraystretch}{1.}
\begin{tabular}{llllllll}
\hline\hline
\multicolumn{8}{l}{Cell-1 (CNN)}                                             \\ \hline
L & Type    & $D$       & $C_\text{in.}$ & $C_\text{out.}$ & $K_s$    & $S_t$    & $g(x)$      \\ \hline
--- & $\text{Input}_1$  & (5,15,10)   & ---       & ---        & ---   & ---   & --- \\
1 & pooling & (5,5,5) & 5         & 5          & (3,2) & (3,2) & ---       \\ 
2 & Conv.   & (16,5,5)   & 5         & 16         & (3,3) & (1,1) & LReLU \\ 
3 & Conv.   & (32,5,5)  & 16        & 32         & (3,3) & (1,1) & LReLU \\ 
4 & Conv.   & (64,5,5)  & 32        & 64         & (3,3) & (1,1) & LReLU \\ 
5 & FC  & 512  & ---       & ---        & ---   & ---   & LReLU \\
6 & FC  & 128   & ---       & ---        & ---   & ---   & LReLU \\
--- & $\text{Out}_1$  & 128   & ---       & ---        & ---   & ---   & --- \\
\hline
\multicolumn{8}{l}{Cell-2(3)(4)(5) (FCNN)}   \\\hline
L & \multicolumn{4}{l}{Type}    &  \multicolumn{2}{l}{$D$}      & \multicolumn{1}{l}{$g(x)$}  \\  \hline
--- & \multicolumn{4}{l}{$\text{Input}_{2(3)(4)(5)}$}  &   \multicolumn{2}{l}{ 4(6)(6)(3)}   & --- \\
1 & \multicolumn{4}{l}{FC}  & \multicolumn{2}{l}{32}      & \multicolumn{1}{l}{LReLU} \\
2 & \multicolumn{4}{l}{FC}  & \multicolumn{2}{l}{64}      & \multicolumn{1}{l}{LReLU} \\
--- & \multicolumn{4}{l}{$\text{Out}_{2(3)(4)(5)}$}  &\multicolumn{2}{l}{64}      & --- \\ \hline
\multicolumn{8}{l}{Cell-6 (FCNN)}   \\\hline
L & \multicolumn{4}{l}{Type}     & \multicolumn{2}{l}{$D$}       & \multicolumn{1}{l}{$g(x)$}  \\  \hline
--- & \multicolumn{4}{l}{$\text{Out}_1 \uplus \text{Out}_2 \uplus ... \uplus \text{Out}_5$}  &   384  & --- \\
1 & \multicolumn{4}{l}{Batch-Norm}  & \multicolumn{2}{l}{384}      & --- \\
2 & \multicolumn{4}{l}{FC}  & \multicolumn{2}{l}{ 512}      & \multicolumn{1}{l}{LReLU} \\
3 & \multicolumn{4}{l}{FC}  & \multicolumn{2}{l}{256}      & \multicolumn{1}{l}{LReLU} \\
4 & \multicolumn{4}{l}{FC}  & \multicolumn{2}{l}{128}      & \multicolumn{1}{l}{Softmax} \\
--- & \multicolumn{4}{l}{Prediction}  & \multicolumn{2}{l}{6}      & --- \\ \hline\hline

\end{tabular}
\end{table}

The COF network consists of 6 cells denoted as Cell-n (n=1-6) as indicated in Table~\ref{tab3}, among which, except for Cell-1, which is a convolutional neural network (CNN), all other cells are fully connected neural networks (FCNN).
Each cell is composed of several unit layers, including convolutional (Conv.), average pooling (pooling), fully connected (FC), and batch normalization (Batch-Norm) layers.

In the presented table, the initial row and the concluding row of each cell respectively signify its input and output. 
In this context, $\text{Input}_1$ comprises five matrices: $C(p_t,y_0)$, $v_1(p_t,y_0)$, $v_2(p_t,y_0)$, $v_3(p_t,y_0)$, $v_4(p_t,y_0)$. 
The $\text{Input}_2$ denotes the mean values of anisotropic flows $\left\langle v_{n} \right\rangle$ ($n=1,2,3,4$) across all emitted particles. 
The $\text{Input}_3$ indicates multiplicities of charged particles, $M_p$, $M_{\pi^-}$, $M_{\pi^+}$, $M_{\Delta^-}$, $M_{\Delta^+}$,  $M_{\Delta^{++}}$.
The $\text{Input}_4$ and $\text{Input}_5$ correspond to the hadron count in transverse and longitudinal emission azimuth angle. 
The input for Cell-6 is a combination of the outputs from Cell-1 to Cell-5 noted as $\text{Out}_1 \uplus \text{Out}_2 \uplus ... \uplus \text{Out}_5$, where the symbol $\uplus$ indicates splicing two vectors, e.g., $[a,b,...] \uplus [c,d,...] = [a,b,...,c,d,...]$.
Besides, the activation function is represented as $\text{LReLU}(x) = \max\{0.01 \times x, x\}$ and $\text{Softmax}(x_i) = {e^{x_i}}/{\sum_j e^{x_j}}$.

Furthermore, to mitigate the influence of absolute data magnitudes, all input features are subject to min-max normalization, ensuring they fall within the range of 0 to 1. 
During the training process, a dynamically decreasing learning rate, which reduces as the loss function converges, is applied with the Adaptive Momentum Estimation (Adam) \cite{Kingma2015.} optimizer.
The aforementioned definitions and concepts are entirely consistent with the patterns used in PyTorch \cite{PyTorchDocs}.

\bibliographystyle{apsrev4-1}
\bibliography{Ref}

\end{document}